\documentclass{aastex63}
\usepackage{graphicx}
\usepackage{gensymb}

\usepackage{graphicx}
\usepackage{wrapfig}
\usepackage{sidecap}
\usepackage{float}
\usepackage{array}
\usepackage{booktabs}
\usepackage{soul}
\shorttitle{Spacecraft exploration of the Uranian satellites}
\shortauthors{Cartwright et al.}

\begin{document}

\title{The science case for spacecraft exploration of the Uranian satellites: \\
		Candidate ocean worlds in an ice giant system}

\correspondingauthor{Richard J. Cartwright}
\email{rcartwright@seti.org}

\author[0000-0002-6886-6009]{Richard J. Cartwright$^a$}
\affiliation{The Carl Sagan Center at the SETI Institute \\
189 Bernardo Avenue, Suite 200\\
Mountain View, CA 94043, USA}
\footnote{$^a$Visiting Astronomer at the Infrared Telescope Facility, which is operated by the University of Hawaii under contract 80HQTR19D0030 with the National Aeronautics and Space Administration. }

\author{Chloe B. Beddingfield}
\affiliation{The Carl Sagan Center at the SETI Institute \\
	189 Bernardo Avenue, Suite 200\\
	Mountain View, CA 94043, USA}
\affiliation{NASA-Ames Research Center\\
	Mail Stop 245-1\\
	Building N245, Room 204\\
	P.O. Box 1\\
	Moffett Field, CA 94035, USA}

\author{Tom A. Nordheim}
\affiliation{Jet Propulsion Laboratory, California Institute of Technology \\
	4800 Oak Grove Drive\\
	Pasadena, CA 91109, USA}

\author{Catherine M. Elder}
\affiliation{Jet Propulsion Laboratory, California Institute of Technology \\
	4800 Oak Grove Drive\\
	Pasadena, CA 91109, USA}

\author{Julie C. Castillo-Rogez}
\affiliation{Jet Propulsion Laboratory, California Institute of Technology \\
	4800 Oak Grove Drive\\
	Pasadena, CA 91109, USA}

\author{Marc Neveu}
\affiliation{NASA Goddard Space Flight Center \\
	8800 Greenbelt Road \\
	Greenbelt, MD 20771, USA}

\author{Ali M. Bramson}
\affiliation{Purdue University \\
	610 Purdue Hall \\
	West Lafayette, IN 47907, USA}

\author{Michael M. Sori}
\affiliation{Purdue University \\
	610 Purdue Hall \\
	West Lafayette, IN 47907, USA}

\author{Bonnie J. Buratti}
\affiliation{Jet Propulsion Laboratory, California Institute of Technology \\
	4800 Oak Grove Drive\\
	Pasadena, CA 91109, USA}

\author{Robert T. Pappalardo}
\affiliation{Jet Propulsion Laboratory, California Institute of Technology \\
	4800 Oak Grove Drive\\
	Pasadena, CA 91109, USA}

\author{Joseph E. Roser}
\affiliation{The Carl Sagan Center at the SETI Institute \\
	189 Bernardo Avenue, Suite 200\\
	Mountain View, CA 94043, USA}
\affiliation{NASA-Ames Research Center\\
	Mail Stop 245-1\\
	Building N245, Room 204\\
	P.O. Box 1\\
	Moffett Field, CA 94035, USA}

\author{Ian J. Cohen}
\affiliation{John Hopkins University Applied Physics Laboratory \\
	11100 John Hopkins Road \\
	Laurel, MD 20723, USA}

\author{Erin J. Leonard}
\affiliation{Jet Propulsion Laboratory, California Institute of Technology \\
	4800 Oak Grove Drive\\
	Pasadena, CA 91109, USA}

\author{Anton I. Ermakov}
\affiliation{University of California, Berkeley \\
	2200 University Avenue \\
	Berkeley, 94720 CA, USA}

\author{Mark R. Showalter}
\affiliation{The Carl Sagan Center at the SETI Institute \\
	189 Bernardo Avenue, Suite 200\\
	Mountain View, CA 94043, USA}

\author{William  M. Grundy}
\affiliation{Lowell Observatory \\
	1400 W Mars Hill Road\\
	Flagstaff, AZ 86001, USA}
\affiliation{Northern Arizona University \\
	S San Francisco Street\\
	Flagstaff, AZ 86011, USA}

\author{Elizabeth P. Turtle}
\affiliation{John Hopkins University Applied Physics Laboratory \\
	11100 John Hopkins Road \\
	Laurel, MD 20723, USA}

\author{Mark D. Hofstadter}
\affiliation{Jet Propulsion Laboratory, California Institute of Technology \\
	4800 Oak Grove Drive\\
	Pasadena, CA 91109, USA}



\begin{abstract}
The 27 satellites of Uranus are enigmatic, with dark surfaces coated by material that could be rich in organics. Voyager 2 imaged the southern hemispheres of Uranus' five largest `classical' moons Miranda, Ariel, Umbriel, Titania, and Oberon, as well as the largest ring moon Puck, but their northern hemispheres were largely unobservable at the time of the flyby and were not imaged. Additionally, no spatially resolved datasets exist for the other 21 known moons, and their surface properties are essentially unknown. Because Voyager 2 was not equipped with a near-infrared mapping spectrometer, our knowledge of the Uranian moons' surface compositions, and the processes that modify them, is limited to disk-integrated datasets collected by ground- and space-based telescopes. Nevertheless, images collected by the Imaging Science System on Voyager 2 and reflectance spectra collected by telescope facilities indicate that the five classical moons are candidate ocean worlds that might currently have, or had, liquid subsurface layers beneath their icy surfaces. To determine whether these moons are ocean worlds, and investigate Uranus' ring moons and irregular satellites, close-up observations and measurements made by instruments onboard a Uranus orbiter are needed. 


\end{abstract} 



\section{Introduction and rationale for a spacecraft mission to Uranus} 
The surfaces of Uranus' large and tidally-locked `classical' moons Miranda, Ariel, Umbriel, Titania, and Oberon exhibit photogeologic evidence for endogenic activity. The ubiquitous evidence hinting at geologic communication between the interiors and surfaces of these moons, in particular on Miranda and Ariel, makes them candidate ocean worlds that may have, or once had, subsurface liquid H$_2$O layers beneath their icy exteriors \citep[]{hendrix2019nasa,beddingfield2020hidden,schenk2020topography}. In 1986, the Voyager 2 spacecraft flew by the Uranian system and collected tantalizing snapshots of these classical moons, measured Uranus' offset and tilted magnetic field, and discovered ten new ring moons orbiting within Uranus' ring system  \citep[e.g.,][]{smith1986voyager}. Since this brief flyby, exploration of Uranus and its satellites has remained in the purview of ground- and space-based telescopes. At near-infrared (NIR) wavelengths, these telescope observations have revealed that the surfaces of the five classical moons are dominated by H$_2$O ice mixed with dark material that could be rich in organics and silicate minerals  \citep[e.g.,][]{cruikshank1977identification,cruikshank1980near,cruikshank1981uranian,soifer1981near,brown1983uranian,brown1984surface}. NIR observations have also detected carbon dioxide (CO$_2$) ice on the classical moons \citep[]{grundy2003discovery,grundy2006distributions,cartwright2015distribution}, and nitrogen-bearing constituents like ammonia (NH$_3$) and ammonium (NH$_4$$^+$) could be present in surface deposits as well \citep[]{bauer2002near,cartwright2018red,cook2018composition,cartwright2020evidence,decolibus2020investigating}. The surfaces of the classical moons could therefore be rich in C, H, O, and N-bearing species, representing some of the key chemical requirements for life as we know it. 

Although data collected by Voyager 2 and ground- and space-based telescope facilities have led to some fascinating discoveries, our understanding of Uranus' classical moons is severely limited by the absence of data collected during close-up observations made by a Uranus orbiter. Similarly, Uranus' small ring moons and irregular satellites remain unexplored, and their surface geologies and compositions are almost entirely unknown.  New measurements made by modern instruments onboard an orbiting spacecraft are critical to investigate the surfaces and interiors of the classical moons and determine whether they are ocean worlds with present day subsurface liquid H$_2$O layers. As described by multiple recent studies \citep[e.g.,][]{fletcher2020ice,cohen2020new,beddingfield2020exploration,leonard2021umami}, a mission to Uranus would allow us to study interactions between Uranus' magnetosphere and its rings and moons, improve our understanding of how geologic processes operate in cold and distant ice giant systems, and enable a more complete investigation of organics in the outer Solar System. An orbiter would provide us with an unparalleled opportunity to study the origin and evolution of Uranus' ring-moon system. Furthermore, an orbiter could provide new insight into the origins of Uranus' irregular satellites, and whether they were sourced from the primordial Kuiper Belt or were captured from nearby heliocentric orbits \citep[e.g.,][]{jewitt2007irregular}. 

Multiple close proximity flybys made by a Uranus orbiter would therefore address many outstanding science questions for the Uranian system (Table 1). A spacecraft mission to Uranus can be carried out with existing chemical propulsion technology by making use of a Jupiter gravity assist in the 2030 -- 2034 time frame, leading to a flight time of only $\sim$11 years, arriving in the early to mid 2040's \citep[]{hofstadter2019uranus}. Crucially, this arrival time frame in northern summer would allow us to observe the Uranian moons' northern hemispheres, which were shrouded by winter darkness at the time of the Voyager 2 flyby and have never been imaged. An orbiter could then continually collect data and search for evidence of ongoing geologic activity, as well as search for evidence of volatile migration of CO$_2$ and other species in response to seasonal changes as the Uranian system transitions into southern spring in 2050.

An orbiter equipped with several key instruments could determine whether the classical Uranian moons are ocean worlds. The highest priority instrument for identifying subsurface oceans is a magnetometer, which could detect and characterize induced magnetic fields emanating from briny liquid layers in the interiors of these moons. Visible (VIS, 0.4 -- 0.7 $\micron$) and mid-infrared (MIR, 5 -- 250 $\micron$) cameras would be vital for identifying recent geologic activity, hot spots, and possible communication between the interiors and surfaces of these moons.  A VIS and NIR mapping spectrometer (0.4 -- 5 $\micron$) would be critical for characterizing the spectral signature and distribution of volatile species that could result from recent outgassing of material, as well as endogenic salts that may have been exposed or emplaced by geologic activity. Ideally, an orbiter would get close enough to achieve spatial resolutions of $\lesssim$ 100 m/pixel for large regions of each classical moon's surface. If only a limited number of close flybys are possible, Ariel and Miranda likely represent the highest priority targets because they display the best evidence for geologic activity in the recent past. 

\begin{table}[h!]
	\centering
	\caption {Science drivers for the exploration of the Uranian system with an orbiter.}
	\hskip-0.8cm\begin{tabular}{c|c|c}
	\toprule
		\begin{tabular}[c]{@{}l@{}}\hspace{-1 cm}Science Questions  \end{tabular} &\begin{tabular}[c]{@{}l@{}}\hspace{-1 cm}Observations and Measurements \end{tabular} & \begin{tabular}[c]{@{}l@{}} \hspace{-1 cm} Instruments\end{tabular}  \\
	\midrule
		\hspace{-0.43 cm}$\bullet$ Do the classical moons have subsurface oceans & $\bullet$ Search for induced magnetic fields, & Magnetometer \\
		\hspace{0.15 cm}that could harbor life, either now or in the past? & plumes, hot spots, and cryovolcanic features. & VIS camera  \\
		\hspace{-0.35 cm}$\bullet$ Is there active geologic communication between &  $\bullet$ Search for surface changes since the & MIR camera \\
		\hspace{-0.77 cm} the classical moons' surfaces and interiors? & Voyager 2 flyby.  & VIS/NIR mapping  \\
		 & $\bullet$ Capture dust samples from & spectrometer  \\
		& active plumes (if present). & Dust spectrometer  \\						
	\midrule
		\hspace{-0.93 cm}$\bullet$ What geologic processes have modified the & $\bullet$ Characterize the morphology, & VIS camera \\
		\hspace{-2.3 cm} surfaces of the classical moons? & 
	    topography, and composition of geologic & VIS/NIR mapping  \\
		& features and terrains on their surfaces. &  spectrometer \\
		& $\bullet$ Measure impact crater densities. &  \\			
	\midrule
		\hspace{0.13 cm}$\bullet$ Do the classical moons have tenuous atmospheres?  & $\bullet$ Search for exospheres.  & VIS camera \\
		\hspace{-0.94 cm} $\bullet$ Do volatile constituents migrate seasonally? & 
		$\bullet$ Characterize the distributions and & VIS/NIR mapping  \\
	   &  spectral signatures of &  spectrometer \\
		& condensed volatiles. &  Plasma spectrometer \\		
		& & UV imaging spectrometer \\
	\midrule
		\hspace{-0.07 cm}$\bullet$ Do magnetospheric charged particles weather the & $\bullet$ Characterize magnetic field and & Magnetometer \\
		\hspace{-.05 cm} surfaces of the ring moons and classical moons?& 
		charged particle populations & Plasma spectrometer  \\
		&  proximal to these moons. &  Energetic particle \\
		& $\bullet$ Search for evidence of &  detector \\
	    & irradiation darkening of surface ices & VIS/NIR spectrometer \\
		& and the presence of radiolytic products. & UV imaging spectrometer \\		
	\midrule
		\hspace{-1.14 cm}$\bullet$ Is the red material on the classical moons & $\bullet$ Make an inbound flyby of an irregular & VIS camera \\
		\hspace{-0.95 cm} organic-rich, and did it originate from the & satellite, and image the irregular & VIS/NIR mapping  \\
	   \hspace{-4.15 cm} irregular satellites? &  satellites while in Uranus orbit. &  spectrometer\\
		& $\bullet$ Characterize the compositions of the &  Dust spectrometer \\	
		& classical moons and collect dust samples. &  \\
	\midrule
		\hspace{-0.68cm}$\bullet$ Does the ring moon Mab sustain the $\mu$-ring?  & $\bullet$ Compare the surface compositions of & VIS camera \\
		\hspace{-0.58 cm} $\bullet$ Does $\mu$-ring material coat Puck and Miranda? & 
		Puck and Miranda to the composition & VIS/NIR mapping \\
		& of Mab and the $\mu$-ring.&  spectrometer \\
		&  $\bullet$ Collect $\mu$-ring dust samples. &  Dust spectrometer \\	
	\midrule
		\hspace{-0.85 cm}$\bullet$ What is the dynamic history of the moons? 
 		& $\bullet$ Measure eccentricities, inclinations, & Radio science \\
		\hspace{-0.28 cm} $\bullet$ Did the classical moons share orbital resonances & 
		tidal Q($\omega$) of Uranus, and Love numbers. & subsystem \\
		\hspace{0.15 cm}in the past, and did these resonances drive tidal & Analyze topography of geologic features &  VIS camera \\	
		 \hspace{-2.28 cm} heating and endogenic activity? &  to estimate paleo heat fluxes. & \\
	\midrule
		 \hspace{-1.33 cm}$\bullet$ Did the classical moons form in Uranus'
		 & $\bullet$ Measure crater frequency distributions. & VIS camera \\
		 \hspace{-0.3 cm} circumplanetary disk or evolve from its rings? &  $\bullet$ Compare surface compositions  & VIS/NIR mapping \\
		 &  of ring moons and classical moons. &  spectrometer \\	
		 &  $\bullet$ Search for streamer channels in the rings. &  \\								
	\bottomrule
	\end{tabular} 
\end{table}

\clearpage

\section{Background and science questions}

Uranus is orbited by 27 known moons (Figure 1), including its five largest moons, Miranda, Ariel, Umbriel, Titania, and Oberon, which have semi-major axes ranging between 5.1 to 23 Uranian radii (R$_U$) (Table 2, Figure 2). Interior to the classical moons, thirteen small ring moons orbit within Uranus’ ring system, with semi-major axes ranging between 2 to 3.8 R$_U$ (Table 3, Figure 3). Far beyond the orbital zone of the classical moons, nine irregular satellites orbit Uranus on highly inclined and eccentric orbits, with semi-major axes ranging between 169 to 805 R$_U$  (Table 3, Figure 1). In the following subsections, we briefly describe the state of knowledge and some of the outstanding science questions for these 27 moons. 

\begin{figure}[h!]
	\centering
	\includegraphics[scale=0.8]{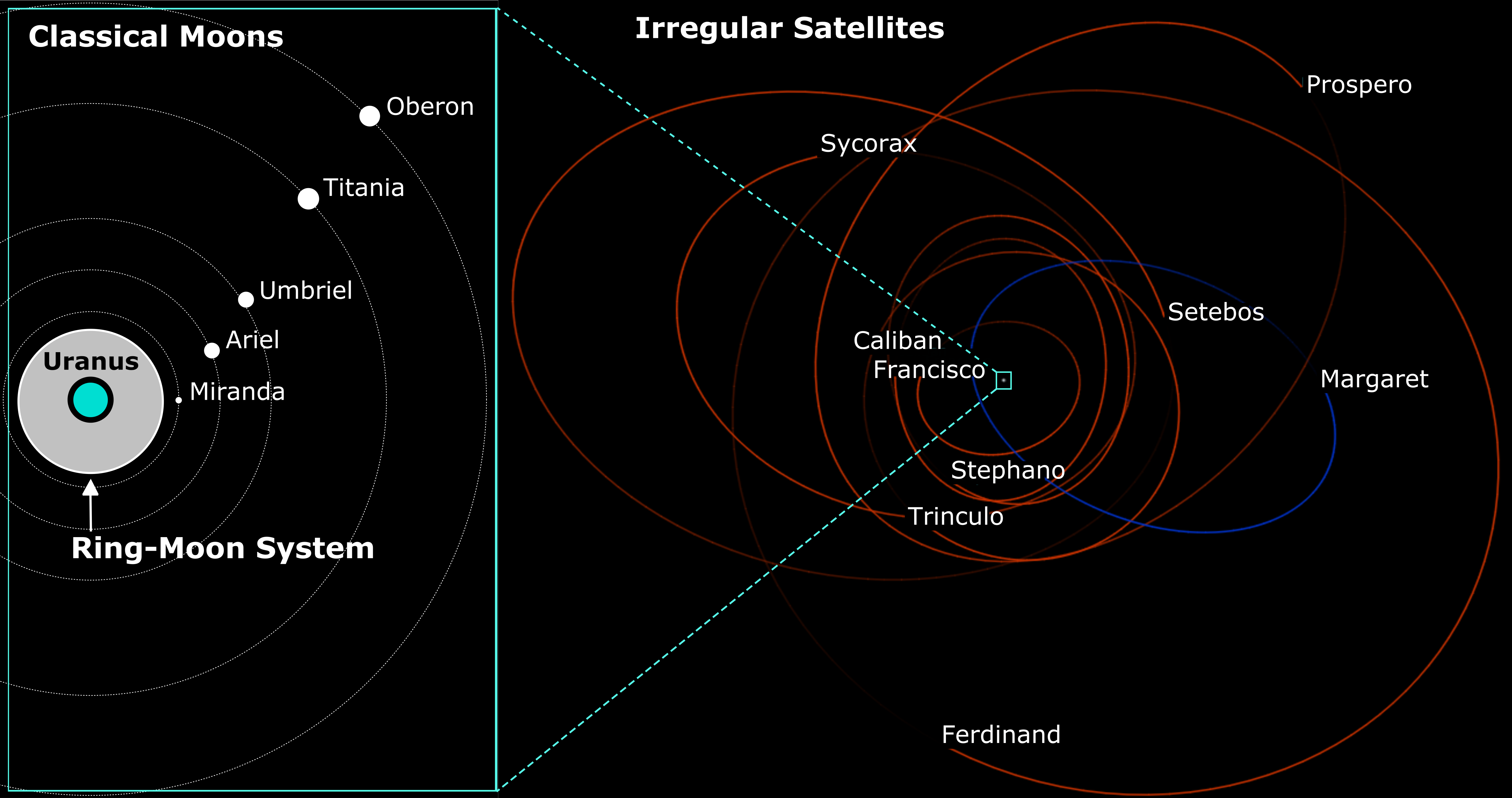}
	\caption{\textit{Left: Diagram illustrating the orbital zones of the ring moons and the classical moons. The sizes of the five classical moons (white filled circles) are scaled correctly relative to each other but have been increased by a factor of 20 relative to the size of Uranus (cyan filled circle). The orbits and relative sizes of the ring moons are shown in Figure 3. Right: Diagram illustrating the orbits of Uranus' eight retrograde irregular satellites (red) and one prograde irregular satellite (blue). The orbital zone of the ring and classical moons is contained within the cyan colored box. The orbits of the irregular satellites were simulated using orbital elements provided on the Jet Propulsion Laboratory Horizons On-Line Ephemeris System \url{(https://ssd.jpl.nasa.gov/?sat_elem}, originally generated by: Nrco0e - Own work, CC BY-SA 4.0, \url{https://commons.wikimedia.org/w/index.php?curid=98634845}).}}\vspace{0.1 cm}
\end{figure} 

\begin{figure}[h!]
	\centering
	\includegraphics[scale=0.44]{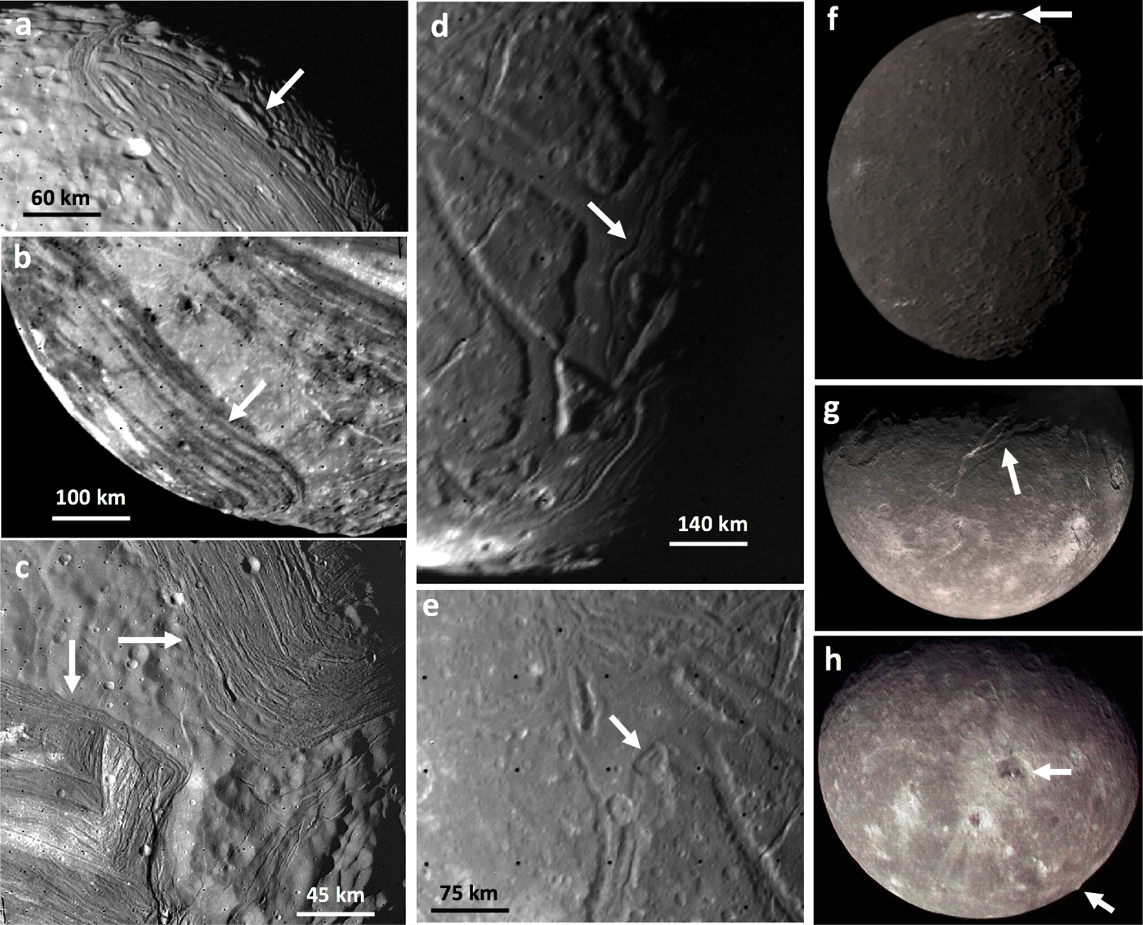}
	\caption{\textit{Voyager 2/ISS images of the five classical Uranian moons. White arrows highlight: (a) ridges on Miranda, which possibly have a cryovolcanic and/or tectonic origin (image shows primarily the trailing hemisphere); (b) Arden Corona on Miranda with high and low albedo banding along large tectonic faults (primarily leading hemisphere); (c) Inverness (bottom left) and Elsinore (top right) Coronae on Miranda that exhibit ridges and grooves (image centered near 45$\degree$S latitude and 300$\degree$ longitude). Between these two coronae are examples of craters that have been mantled by an unknown source of material. (d) Large chasmata with medial grooves on Ariel (primarily trailing hemisphere); (e) an impact crater on Ariel, possibly infilled by cryolava (centered near 50$\degree$S latitude, 10$\degree$ longitude). (f) The bright floor of Wunda crater on Umbriel; (g) the large Messina Chasmata on Titania; and (h) the smooth floor of Hamlet crater and an 11 km tall ‘limb mountain’ on Oberon. The ISS image mosaics in f-h show both the leading and trailing hemispheres of these three moons, with their south poles located toward the bottom center of each image (reprocessed by \citealt{stryk2008voyager}). A grid of Reseau points were embedded in images collected by ISS (regular pattern of black-filled circles in a-e).}}\vspace{0.1 cm}
\end{figure} 

\begin{table}[tbp]
	\centering
	\caption {Overview of the five classical moons' geologies and surface compositions.}
	\hskip-0.8cm\begin{tabular}{*6c|c}
		\toprule
		\begin{tabular}[c]{@{}l@{}}\hspace{-1 cm}Classical \\ \hspace{-1 cm}Moons  \end{tabular} &\begin{tabular}[c]{@{}l@{}}\hspace{-1 cm}*Semi-major\\  \hspace{-1 cm}Axis (10$^5$ km) \end{tabular} & \begin{tabular}[c]{@{}l@{}} \hspace{-1 cm}*Mean \\  \hspace{-1 cm}Radius (km)\end{tabular} & \begin{tabular}[c]{@{}l@{}} \hspace{-1 cm}Geologic Features \\  \hspace{-1 cm} and Regions of Interest \end{tabular} &  \begin{tabular}[c]{@{}l@{}} \hspace{-1 cm}Highest Res. \\  \hspace{-1 cm}Images (km/pix)\end{tabular} & \begin{tabular}[c]{@{}l@{}} \hspace{-1 cm}Known Surface \\  \hspace{-1 cm}Constituents \end{tabular} &\begin{tabular}[c]{@{}l@{}} \hspace{-1 cm}Possible Surface \\ \hspace{-1 cm} Constituents \end{tabular}  \\
		\midrule	
		Miranda & 1.299 & 235.8 $\pm$ 0.7   & Polygonal shaped coronae, &  $\sim$0.25 &H$_2$O ice & \\
		&  &  &   large scale rift system, & &  &  \\
		&  &  &   populations of subdued  & &  &  \\
		&  &  &   and fresh craters. & &  & \\ [0.1cm]
		Ariel & 1.909 & 578.9 $\pm$ 0.6 & Large chasmata, & $\sim$1 &  H$_2$O ice &  \\
		&  &  &  flow bands, smooth plains, & & CO$_2$ ice  & $^\dagger$Organics,   \\	
		&  &  & cratered plains, large & &  & $^\dagger$hydrated silicates,  \\	
		&  &  &  lobate shaped features. & &  & $^\dagger$$^\dagger$NH$_3$ and NH$_4$  \\ [0.1cm]
		Umbriel & 2.660 & 584.7 $\pm$ 2.8 & Heavily cratered surface & $\sim$5 &  H$_2$O ice & bearing species,   \\
		&  &  & with some bright-floored & & CO$_2$ ice  & $^\dagger$$^\dagger$carbonates,  \\	
		&  &  &  craters, polygonal basins. & &  & $^\dagger$$^\dagger$$^\dagger$trapped OH  \\ [0.1cm]	
		Titania & 4.363 & 788.9 $\pm$ 1.8 & Large chasmata, smooth & $\sim$3 & H$_2$O ice  \\
		&  &  &  plains, linear features,  & & CO$_2$ ice  &   \\		
		&  &  & cratered terrain. & &  &  \\ [0.1cm]
		Oberon & 5.835 & 761.4 $\pm$ 2.6 &  Large chasmata, smooth &  $\sim$6 & H$_2$O ice &   \\
		&  &  & plains, cratered terrain, & & CO$_2$ ice  &  \\	
		&  &  & $\sim$11 km tall mountain. &  & &  \\		
		\bottomrule
	\end{tabular} 
	\textit{*Semi-major axis and mean radius values from the Jet Propulsion Laboratory Horizons On-Line Ephemeris System, and \\
		\hspace{-9.4 cm} references therein \url{(https://ssd.jpl.nasa.gov/?sat_elem).}} \\
	\hspace{-2.48 cm} \textit{$^\dagger$Possibly contributing to the ubiquitous dark material and spectrally red material observed by Voyager 2. }\\
	\hspace{-4.35 cm} \textit{$^\dagger$$^\dagger$Possible contributors to a 2.2-$\micron$ absorption band detected in some ground-based spectra.} \\
	\hspace{-3.4 cm} \textit{$^\dagger$$^\dagger$$^\dagger$Possibly contributing to a 0.28-$\micron$ absorption band detected in Hubble Space Telescope spectra.} \\	
\end{table}

\subsection{Geology of the Uranian Satellites}
\textit{Classical Moons:} The Imaging Science System (ISS) onboard Voyager 2 collected fascinating images of the five classical moons (Figure 2). Because the Voyager 2 flyby occurred during southern  summer, when the subsolar point was $\sim$81$\degree$S, the northern hemispheres of these moons were mostly hidden from view and were not imaged \citep[]{smith1986voyager}. The incomplete spatial coverage, and generally low spatial resolution of the available images, limits our understanding of different terrains and geologic features, in particular for the more distant moons Umbriel, Titania, and Oberon. 

The innermost moon Miranda displays abundant evidence for endogenic geologic activity, including three large polygonal shaped regions called coronae, which were likely formed by tectonic and/or cryovolcanic processes (Figure 1a-c) \citep[e.g.,][]{smith1986voyager,croft1991geology,greenberg1991miranda,schenk1991fluid,kargel1995cryovolcanism,pappalardo1997extensional,hammond2014global,beddingfield2015fault,beddingfield2020hidden}. The origin and time scale of activity on Miranda is not well understood, and it is unknown if this activity is associated with a subsurface ocean, either now or in the past. Searching for and characterizing induced magnetic fields, plumes, and surface heat anomalies, as well as analyzing geologic surface features interpreted to be cryovolcanic in origin, is paramount to determine if Miranda is an ocean world. Tidal heating of Miranda from past orbital resonances \citep{tittemore1990tidal,cuk2020dynamical} may have been an important driver of resurfacing in the recent past. Additionally, Miranda displays ancient cratered terrain pockmarked with `subdued' craters, which have smooth floors and subtle rims that have been mantled by an unknown source of material \citep[e.g.,] []{smith1986voyager,beddingfield2020hidden,cartwright2020regolith}. These subdued craters are reminiscent of lunar highlands and could be blanketed by impact-generated regolith \citep[]{croft1987miranda}, but they are also reminiscent of the plume-mantled craters on the ocean world Enceladus \citep[e.g.,][]{kirchoff2009crater}, hinting that a similar plume-driven mantling process may have occurred on Miranda. 

Miranda’s neighbor Ariel also displays widespread evidence for resurfacing that could have been spurred by ocean world activity. Sections of Ariel's imaged surface are dominated by large canyons called `chasmata' (Figure 2d, 2e) that were likely formed by tectonic processes \citep[e.g.,][]{smith1986voyager}. The smooth floors of some of these chasmata are bowed up with two parallel medial ridges that are separated by a topographic low, reminiscent of fissure style volcanism on Earth \citep[]{schenk1991fluid}. Large fracture systems cut across other parts of Ariel’s surface, and clusters of curvilinear features referred to as `flow bands' could be cryovolcanic features \citep[e.g.,][]{plescia1987geological,croft1991geology,schenk1991fluid,beddingfield2021Arielcryo}. Furthermore, Ariel's surface geology is consistent with high heat fluxes in its interior, which could drive extrusion of material \citep[]{peterson2015elastic}. Regions of Ariel’s surface are relatively young ($\sim$1--2 Ga) \citep{zahnle2003cratering}, but the fairly low resolution of the ISS images ($\gtrsim$ 1 km/pixel) makes investigation of the processes that formed these younger terrains more difficult. 

Although Umbriel has the darkest and oldest surface of the five classical moons \citep[e.g.][]{smith1986voyager,zahnle2003cratering}, it has some large craters with bright floors, like Wunda crater, which has a bright annulus of material surrounding its central peak \citep[e.g.,][]{smith1986voyager} (Figure 2f). Wunda crater is located near the center of Umbriel's trailing hemisphere, where the abundance of CO$_2$ ice is highest \citep[]{grundy2006distributions,cartwright2015distribution}. Consequently, the bright annulus of material mantling this crater may represent a large deposit of CO$_2$ ice, possibly originating from post-impact cryovolcanic infilling \citep[]{helfenstein1989evidence}, or from the accumulation of radiolytically-produced CO$_2$ molecules that get cold trapped in Wunda \citep[]{sori2017wunda}. Furthermore, large polygonal basins are present on Umbriel, hinting at global-scale resurfacing in the past \citep[]{helfenstein1989evidence}. However, the poor resolution of the available data ($\gtrsim$ 5 km/pixel) severely limits analyses of these features and our ability to determine whether they resulted from ocean world activity. 

The surfaces of the outer moons Titania and Oberon exhibit evidence for tectonism, with large chasmata and linear surface features, as well as smooth plains that may have formed from cryovolcanic processes \citep[e.g.,][]{smith1986voyager,croft1991geology} (Figure 2g, 2h). Additionally, Oberon has an $\sim$11 km tall `limb mountain' that could be the central peak for a relaxed complex crater \citep[]{smith1986voyager,mckinnon1991cratering,croft1991geology,moore2004large}. Similar to Umbriel, the poor resolution of the ISS images for Titania and Oberon ($\gtrsim$ 3 and 6 km/pixel, respectively) limits our ability to investigate possible ocean world activity on these two moons. 

Relatively less is known about the interiors of the classical moons, and measuring magnetic induction with a magnetometer is critical for determining whether they possess subsurface oceans \citep[]{cochrane2021induced,weiss2021searching}. Other measurements are needed to investigate the internal structures and bulk compositions of these moons, which is critical for improving our understanding of their formation and evolution. The densities of Ariel, Umbriel, Titania, and Oberon are 1.66 $\pm$ 0.15 g cm$^-$$^3$, 1.39 $\pm$ 0.16 g cm$^-$$^3$, 1.71 $\pm$ 0.05 g cm$^-$$^3$, and 1.63 $\pm$ 0.05 g cm$^-$$^3$, respectively, indicating that these moons are made of at least 50$\%$ silicate material by mass, whereas Miranda's density is only 1.2 $\pm$ 0.14 g cm$^-$$^3$, indicating a larger H$_2$O ice fraction \citep[]{jacobson1992masses}. Measuring the non-spherical gravity field would shed light on the differentiation state and the nature of the endogenic activity exhibited by these moons. Measuring libration amplitudes \citep[]{steinbrugge2019measuring} could also be used to investigate whether these moons have subsurface liquid layers. Furthermore, searching for thermal anomalies across the surfaces of these moons, using a MIR camera equipped with a suite of multiple narrow and wide filters spanning 5 to 250 $\micron$, would improve our understanding of their heat budgets and the long-term survivability of liquid water in their interiors. Thus, characterizing the surfaces and interiors of the classical moons to determine whether they are, or were, ocean worlds requires high resolution datasets, which can only be collected by a Uranus orbiter making multiple close flybys of each moon. 

\begin{figure}[h!]
	\centering
	\includegraphics[scale=0.4]{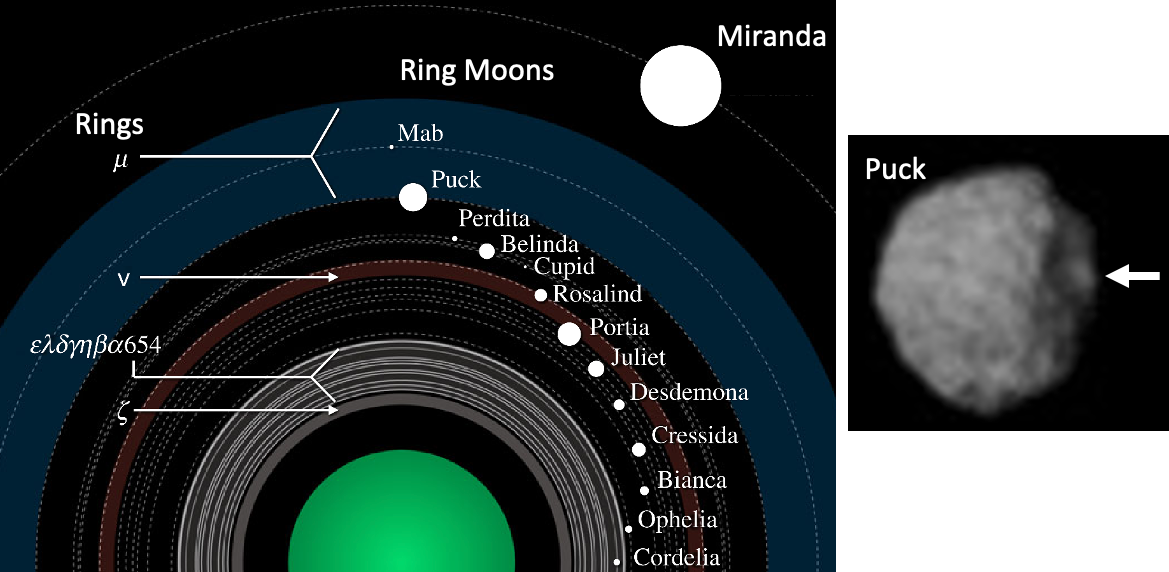}
	\caption{\textit{Left: Diagram illustrating Uranus' densely packed ring and ring-moon system (modified from \citealt{showalter2020rings}). The sizes of the moons (white filled circles) are scaled correctly relative to each other but have been increased by a factor of 40 relative to the size of Uranus (green filled circle). Right: Voyager 2/ISS image of Puck (reprocessed by \citealt{stryk2008voyager}). The white arrow highlights Bogle crater, the largest crater on Puck ($\sim$47 km diameter). Puck's south pole is located near the center of the left-side of the image. }}\vspace{0.1 cm}
\end{figure} 

\textit{Ring Moons and Irregular Satellites:} Uranus possesses the most densely packed group of moons in the Solar System, with nine ring moons (Bianca to Perdita) orbiting between 59,000 to 77,000 km from the planet's center \citep[e.g.,][]{showalter2020rings}. Voyager 2 initially discovered ten ring moons: Cordelia, Ophelia, Bianca, Cressida, Desdemona, Juliet, Portia, Rosalind, Belinda, and Puck \citep[e.g.,][]{smith1986voyager}. Perdita was discovered later through reanalysis of Voyager 2 images \citep{karkoschka2001comprehensive}. Cupid and Mab were discovered using the Hubble Space Telescope \citep[]{showalter2006second,de2006new}. Mab orbits within the outermost and dusty $\mu$-ring (blue colored region in Figure 3), which might be sustained by material ejected from the surface of this tiny moon \citep[]{showalter2006second,de2006new,sfair2012role}. Furthermore, $\mu$-ring material could spiral inward and coat the leading hemisphere of Puck \citep[]{french2017orbital}, and perhaps it also spirals outward and mantles Miranda, possibly contributing to its substantial regolith cover \citep[]{cartwright2020regolith,cartwright2020probing}. Little is known about the surface geologies of the ring moons as only Puck was spatially resolved by Voyager 2/ISS ($\sim$4.5 km/pixel) (Figure 3). These ISS images revealed a heavily cratered surface, suggesting that Puck may have been collisionally disrupted and then reaccreted into a rubble pile \citep[]{smith1986voyager,croft1991geology}.  Voyager 2 did not detect any of Uranus' nine known irregular satellites, which were discovered by ground-based observers \citep[]{gladman1998discovery,gladman2000discovery,kavelaars2004discovery,sheppard2005ultradeep}. Thus, the geologies of Uranus’ ring moons and irregular satellites remain unexplored, and new observations made by an orbiter would dramatically expand our knowledge of these objects. 

\setlength{\abovetopsep}{-3pt} 

\begin{table}[h!]
	\centering
	\caption {Ring moons and irregular satellites of Uranus.}
	\hskip-0.8cm\begin{tabular}{*3c|c}
		\toprule
		\begin{tabular}[c]{@{}l@{}}\hspace{-1 cm}Satellites  \end{tabular} &\begin{tabular}[c]{@{}l@{}}\hspace{-1 cm}*Semi-major\\  \hspace{-1 cm}Axis (10$^5$ km) \end{tabular} & \begin{tabular}[c]{@{}l@{}} \hspace{-1 cm}*Mean \\  \hspace{-1 cm}Radius (km)\end{tabular} & \begin{tabular}[c]{@{}l@{}} \hspace{-1 cm}Possible Surface \\ \hspace{-1 cm} Constituents \end{tabular}  \\
		\midrule
		\textit{Ring Moons}  &  &  &   \\
		Cordelia & 0.4980 & 20.1 $\pm$ 3 &    \\
		Ophelia & 0.5380 & 21.4 $\pm$ 4  &   \\
		Bianca & 0.5920 & 27 $\pm$ 2  &   \\
		Cressida  & 0.6180 & 41 $\pm$ 2 &    \\
		Desdemona  & 0.6270 & 35 $\pm$ 4  &  \\
		Juliet  & 0.6440 & 53 $\pm$ 4  & H$_2$O ice, \\
		Portia & 0.6610 & 70 $\pm$ 4  &   $^\dagger$organics,   \\
		Rosalind  & 0.6990 & 36 $\pm$ 6 &  $^\dagger$hydrated silicates   \\
		Cupid  & 0.7439 & 9 $\pm$ 1  &   \\
		Belinda  & 0.7530 & 45 $\pm$ 8  &  \\
		Perdita & 0.7642 & 13 $\pm$ 1 &   \\
		Puck & 0.8600 & 81 $\pm$ 2 &   \\
		Mab & 0.9774 & 6 -- 12 &    \\	
		\midrule	
		\textit{Irregular Satellites} &  &  &   \\
		Francisco & 42.83 & 11 &  \\
		Caliban & 72.31 & 36 &    \\
		Stephano & 80.07 & 16 & \\
		Trinculo & 85.05 & 9 &   H$_2$O ice,  \\
		Sycorax & 121.8 & 75 &   $^\dagger$organics, \\
		Margaret & 141.5 & 10 & $^\dagger$hydrated silicates  \\
		Prospero & 162.8 & 25 &   \\
		Setebos & 174.2 & 24 &    \\
		Ferdinand & 204.3 & 10 &   \\
		\bottomrule
	\end{tabular} 
	
	\textit{*Semi-major axis and mean radius values from the Jet Propulsion \\
		\hspace{-0.3 cm}Laboratory Horizons On-Line Ephemeris System, and references \\ 
		\hspace{0.3 cm}therein \url{(https://ssd.jpl.nasa.gov/?sat_elem)}. Radius range for Mab \\
		\hspace{-3.85 cm}reported in \citealt{showalter2006second}.} \\
	\hspace{0.1 cm} \textit{$^\dagger$Possibly contributing to the ubiquitous dark material, as well as the \\
		\hspace{-1.4 cm} spectrally red material detected on the irregular satellites. }\\
\end{table}

\vspace{0 cm}\subsection{Surface Compositions of the Uranian Satellites}

\textit{Classical Moons:} Ground-based, NIR telescope observations ($\sim$1 -- 2.5 $\micron$) determined that the classical moons have surface compositions dominated by a mixture of H$_2$O ice and a dark, spectrally neutral material of unknown origin \citep[e.g.,][]{cruikshank1977identification,brown1983uranian,brown1984surface}. Laboratory experiments indicate that the dark material has a spectral signature similar to amorphous carbon and/or silicates \citep[]{clark1984spectral}. More recent NIR telescope observations have determined that the Uranian moons display leading/trailing and planetocentric asymmetries in their compositions (Figure 4). For example, `pure' CO$_2$ ice (i.e., segregated from other constituents in concentrated deposits) has been detected on the trailing hemispheres of Ariel, Umbriel, Titania, and Oberon, with larger abundances on the inner moons, Ariel and Umbriel, compared to the outer moons, Titania and Oberon (Figure 5). \citep[]{grundy2003discovery,grundy2006distributions,cartwright2015distribution}. CO$_2$ ice on these moons could be generated via irradiation of native H$_2$O ice and C-rich material by magnetospheric charged particles \citep[]{grundy2006distributions,cartwright2015distribution}. Supporting this hypothesis, trapped OH has possibly been detected on these moons \citep[]{roush1997uv}, which is a strong catalyst for radiolytic generation of CO$_2$ molecules from substrates composed of H$_2$O ice and carbon-rich species \citep[e.g.,][]{mennella2004formation,raut2012radiation}.  

\begin{figure}[h!]
	\centering
	\includegraphics[scale=0.5]{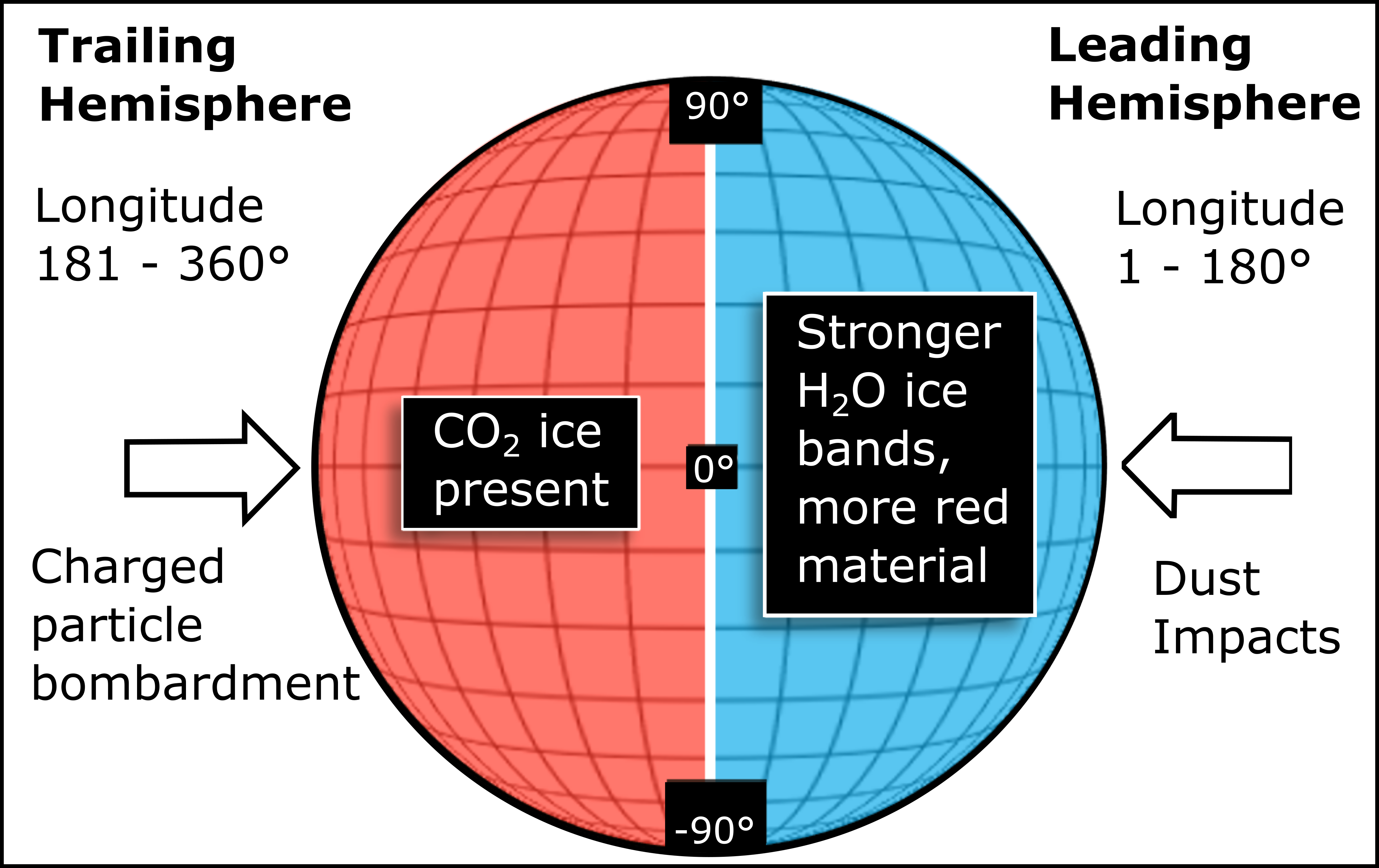}
	\caption{\textit{Diagram showing the broad leading/trailing hemispherical trends in composition exhibited by Ariel, Umbriel, Titania, and Oberon, possibly driven by charged particle interactions (primarily with their trailing hemispheres) and heliocentric and planetocentric dust impacts (primarily with their leading hemispheres).}} \vspace{0.1 cm}
\end{figure} 

\begin{figure}[h!]
	\centering
	\includegraphics[scale=0.6]{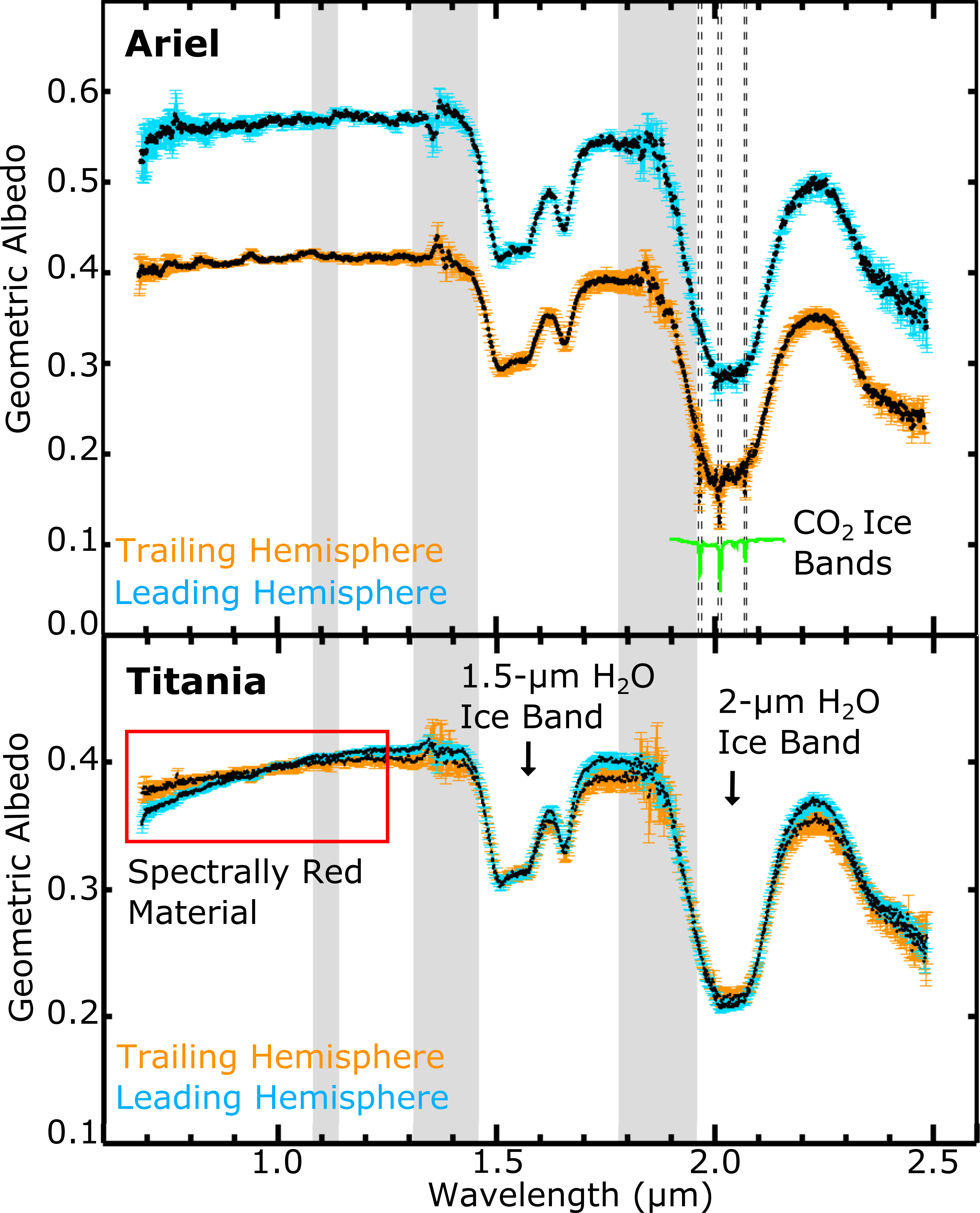}
	\caption{\textit{Top: Example NIR spectra of Ariel's leading and trailing hemisphere (blue and orange 1$\sigma$ error bars, respectively), collected with the SpeX Spectrograph on NASA's Infrared Telescope Facility (IRTF) \citep{rayner2003spex}, offset vertically for clarity. These data demonstrate the strong leading/trailing longitudinal asymmetry in the distribution of CO$_2$ ice observed on these moons (wavelength positions of prominent CO$_2$ bands are indicated with black dashed lines). The observed CO$_2$ ice bands are compared to an example laboratory spectrum of `pure' CO$_2$ ice (green) \citep{hansen1997spectral}. Bottom: Example IRTF/SpeX spectra of Titania's leading and trailing hemisphere (blue and orange 1$\sigma$ error bars, respectively). These data highlight the NIR wavelength region where spectrally red material has been detected on these moons (0.7 -- 1.25 $\micron$, red box), with more red material on their leading hemispheres. These data also show the leading/trailing longitudinal asymmetry in H$_2$O ice band strengths, which are stronger on the leading hemispheres of Ariel, Umbriel, Titania, and Oberon. Gray-toned zones represent wavelength regions with lower atmospheric transparency and increased telluric contamination.}} \vspace{0.1 cm}
\end{figure} 

H$_2$O ice bands are weaker on the trailing hemispheres of these moons  (Figure 5), perhaps in part due to large deposits of CO$_2$ ice masking the NIR spectral signature ($\sim$1 -- 2.5 $\micron$) of H$_2$O ice \citep[]{cartwright2015distribution,cartwright2020probing}. Another possibility is that heliocentric dust impacts promote regolith overturn and expose ‘fresh’ H$_2$O ice, primarily on these moons' leading hemispheres \citep[]{cartwright2018red}. Spectrally red material ($\sim$0.4 -- 1.3 $\micron$) has also been detected, primarily on the leading hemispheres of the outer moons, Titania and Oberon  (Figure 5) \citep[]{buratti1991comparative,bell1991search,helfenstein1991oberon,cartwright2018red}. The distribution of red material could result from the accumulation of infalling planetocentric dust from retrograde irregular satellites \citep[]{buratti1991comparative,tamayo2013chaotic,cartwright2018red}, which are spectrally redder than the classical moons \citep[e.g.,][]{grav2004photometry,maris2007light}. 

Over longer wavelengths ($\sim$3 -- 6.5 $\micron$), spectrophotometric data collected by the Infrared Array Camera \citep{fazio2004infrared} on the Spitzer Space Telescope \citep{werner2004spitzer} indicate that the spectral signature of CO$_2$ ice is strangely absent from all of these moons \citep[]{cartwright2015distribution,cartwright2020probing}. One possible explanation is that the regoliths of the classical moons are mantled by a thin layer of small H$_2$O ice grains ($\le$ 2 $\micron$ diameters), which obscures the longer wavelength spectral signature of CO$_2$ ice retained beneath this topmost layer \citep[]{cartwright2015distribution,cartwright2018red,cartwright2020probing}.  Supporting this possibility, visible wavelength polarimetry data suggest that these moons have porous regoliths, dominated by small grains \citep[]{afanasiev2014polarimetry}. Similarly, over longer mid-infrared wavelengths (20 -- 50 $\micron$), data collected by the Infrared Interferometer Spectrometer (IRIS) on Voyager 2 suggest that Miranda and Ariel have regoliths with unusual microstructures, possibly dominated by dark, isotropically scattering grains \citep{hanel1986infrared}. These different datasets all suggest that the regoliths of the Uranian moons are notably different from both H$_2$O ice-rich and dark material-rich Galilean and Saturnian moons and may be more comparable to H$_2$O ice-rich trans-Neptunian objects \citep[]{afanasiev2014polarimetry,cartwright2020probing,detre2020herschel}. 

Although these results are intriguing, new spacecraft measurements are needed to better understand the processes modifying the surface compositions of the Uranian moons. For example, the distribution of CO$_2$ ice is only longitudinally constrained, limiting our ability to determine whether this volatile is generated by charged particle radiolysis, or whether it is a native constituent sourced from their interiors. Unlike the other classical moons, CO$_2$ ice and leading/trailing longitudinal asymmetries in composition are absent on Miranda \citep[]{bauer2002near,grundy2006distributions,gourgeot2014near,cartwright2018red,cartwright2020probing,decolibus2020investigating}, adding to the mystery surrounding this moon. Because of the large obliquity of the Uranian system, and the associated seasonal effects, CO$_2$ ice exposed or generated at polar latitudes on these moons should sublimate, migrate in tenuous exospheres to their low latitudes, and condense in cold traps \citep[]{grundy2006distributions,sori2017wunda,cartwright2021arielCO2}. Spatially resolved spectra collected by an orbiter are needed to more completely understand the origin and nature of CO$_2$ ice. Additionally, the regolith properties of the classical moons could result from interactions with the surrounding space environment, which cannot be properly assessed without data collected by an orbiter.  

Some ground-based spectra of the classical moons show a 2.2-$\micron$ absorption band (Figure 6) \citep[]{bauer2002near,cartwright2018red,cartwright2020evidence}, which is similar to a 2.2-$\micron$ feature attributed to NH$_3$ and NH$_4$-bearing species on Pluto and its moons \citep[e.g.,][]{brown2000evidence,grundy2016surface,cook2018composition,dalle2019detection,cruikshank2019recent,protopapa2020charon} and the Saturnian moon Enceladus \citep[]{emery2005near,verbiscer2006near}. NH$_3$ and NH$_4$ are highly efficient anti-freeze agents when mixed with liquid H$_2$O that could promote the retention of subsurface oceans if present within the interiors of these moons \citep[e.g.,][]{spohn2003oceans,hussmann2015interiors,nimmo2016ocean}. Because NH$_3$-rich planetesimals were likely incorporated into the proto-Uranian moons as they formed in the Uranian subnebula \citep[e.g.,][]{lewis1972metal}, large quantities of NH$_3$ could be present in their interiors. Furthermore, unlike H$_2$O ice and CO$_2$ ice, the 2.2-$\micron$ band does not display leading/trailing longitudinal asymmetries in its distribution (Figure 6), which suggests that NH$_3$-bearing and NH$_4$-bearing species are native to these moons and are exposed by impact events, tectonism, and mass wasting, and/or emplaced in cryovolcanic deposits \citep[]{cartwright2020evidence}. Other species like carbonates (-CO$_3$ group), including ammonium carbonate ((NH$_4$)$_2$CO$_3$), could be contributing to the 2.2-$\micron$ band as well  \citep[]{cartwright2020evidence}. Higher spatial resolution spectra are needed to measure the spectral signature and spatial distribution of the 2.2-$\micron$ band to determine whether these moons are rich in NH$_3$ and other species that may have been sourced from liquid layers in their interiors. 

\begin{figure}[h!]
	\centering
	\includegraphics[scale=0.29]{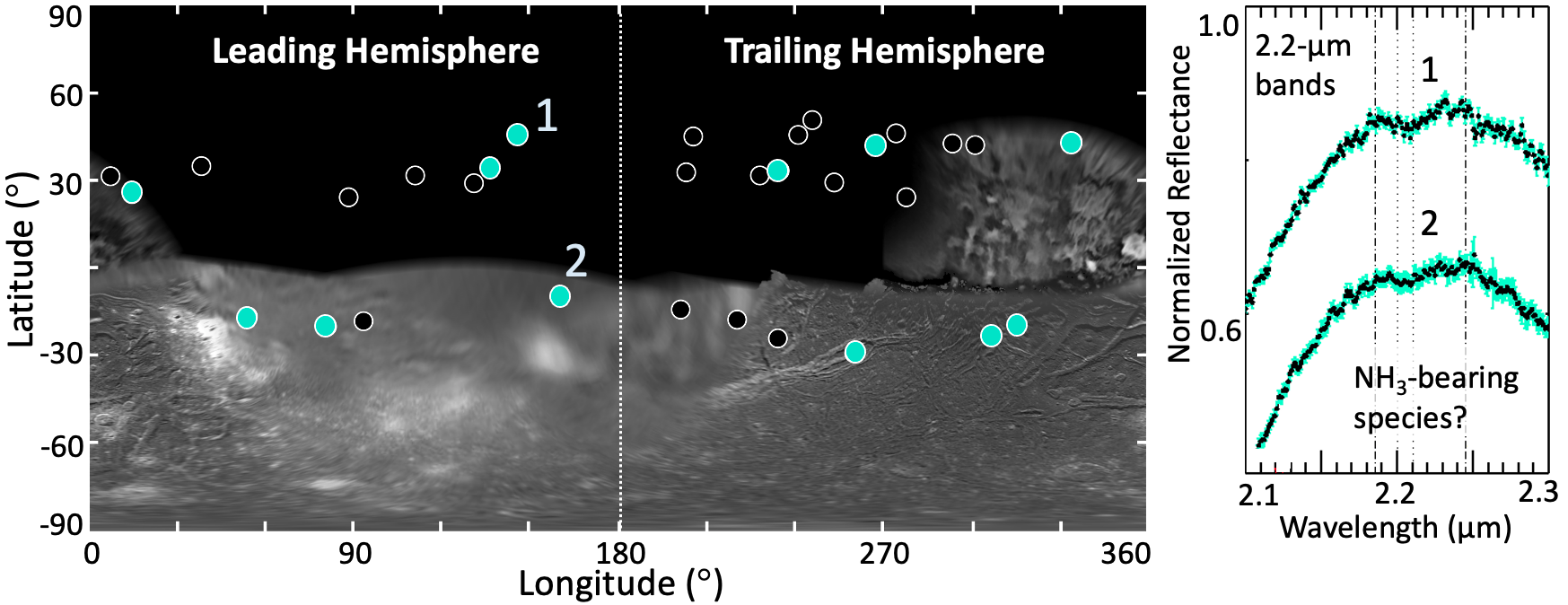}
	\caption{\textit{Left: Mercator map projection of Voyager 2/ISS images of Ariel (courtesy NASA/JPL/Caltech/USGS, reprocessed by \citealt{stryk2008voyager}).The cyan dots represent the mid-observation longitude and latitude for ground-based spectra that show 2.2-$\micron$ bands, and the black dots represent spectra that do not display 2.2-$\micron$ bands (modified from \citealt{cartwright2020evidence}). These spectra are disk-integrated and average over an entire hemisphere of Ariel. Spectra for the dots labeled `1' and `2' are shown on the right. Right: Example IRTF/SpeX spectra of Ariel that display 2.2-$\micron$ bands, offset vertically for clarity. Dot-dashed lines highlight the wavelength range of the 2.2-$\micron$ band, and the dotted lines highlight the wavelength range of its band center.}}\vspace{0.1 cm}
\end{figure} 

The composition and origin of the widespread dark material and the spectrally red material on the Uranian moons remains poorly understood. These materials could be rich in hydrated silicates and organic constituents delivered to and/or native to these moons. Prior spacecraft missions have assessed the nature of organics in the Jupiter, Saturn, and Pluto systems, as well as on comets \citep[e.g.,][]{mccord1997organics,irvine2003hcn,clark2005compositional,waite2009liquid,capaccioni2015organic,grundy2016surface,cruikshank2020organic}. Measuring the spectral signature of organics (e.g., C-H stretching modes between 3.2 and 3.5 $\micron$) in the Uranian system represents a key heliocentric link for improving our understanding of the nature and overall distribution of organic matter in the Solar System, as well as for investigating whether organics formed within the protoplanetary disk or were delivered as interstellar matter. Thus, new measurements made by an orbiter are critical for determining the spatial distribution and spectral signature of CO$_2$ ice and NH$_3$-bearing species on the classical moons, and whether these constituents were exposed/emplaced on their surfaces by ocean world activity, as well as for investigating the origin and evolution of organic material in the Uranian System. 

\textit{Ring Moons and Irregular Satellites:} Far less is known about the compositions of Uranus’ smaller ring moons and irregular satellites, which are too faint for spectroscopic observations using available telescopes (apparent magnitudes ranging between 19.8 and 25.8 at visible wavelengths). Photometric datasets indicate that the ring moons are dark (except possibly Mab, \citealt[]{showalter2006second}), with neutral spectral slopes and slight reductions in albedo at 1.5 and 2.0 $\micron$, consistent with the presence of H$_2$O ice \citep[e.g.,][]{karkoschka2001comprehensive}. More recent photometric measurements indicate that the ring moons may display latitudinal variations in albedo, possibly resulting from interactions with Uranus' magnetosphere \citep[]{paradis2019photometry}. Other photometric studies determined that Uranus’ irregular satellites are dark and red \citep[e.g.,][]{grav2004photometry,maris2007light}, with possibly redder colors than the irregular satellites of the other giant planets \citep[]{graykowski2018colors}. Spectra of the largest Uranian irregular satellite Sycorax suggest that H$_2$O ice is present \citep[]{romon2001photometric}, but no spectra exist for the other, fainter irregular satellites. Therefore, the compositions of these objects are essentially unknown, and new observations made by an orbiter are needed.

\section{Conclusions and Recommendations for Future Exploration} 
Data returned by Voyager 2 and ground- and space-based telescopes have revealed tantalizing glimpses of the Uranian moons' geologic histories, compositions, and interactions with the surrounding space environment. However, Uranus' 27 satellites remain poorly understood. Only six Uranian moons were spatially resolved by ISS, and the flyby nature of the Voyager 2 encounter, the lack of a mapping spectrometer, and the low spatial resolution of collected datasets left many unanswered questions and limit our ability to determine whether the classical moons are, or were, ocean worlds. Future telescope facilities like the Extremely Large Telescopes (ELTs) and the proposed Large UV/Optical/IR Surveyor (LUVOIR) space telescope \citep[e.g.,][]{roberge2019telling} will be able to collect high quality images and spectra of the classical moons \citep[]{cartwright2019exploring,wong2020transformative}, providing new information about their surfaces (Figure 7). Although these telescope datasets will undoubtedly increase our knowledge of these moons, they will not be able to achieve sufficient spatial resolutions to identify linkages between geologic features and their surface compositions, nor probe their internal structures, or investigate moon-magnetosphere interactions \citep[e.g.,][]{kollmann2020magnetospheric}. Furthermore, these telescope datasets will not be able to spatially resolve the ring moons and irregular satellites to assess their surface geologies and origins. 

\begin{figure}[h!]
	\centering
	\includegraphics[scale=0.3]{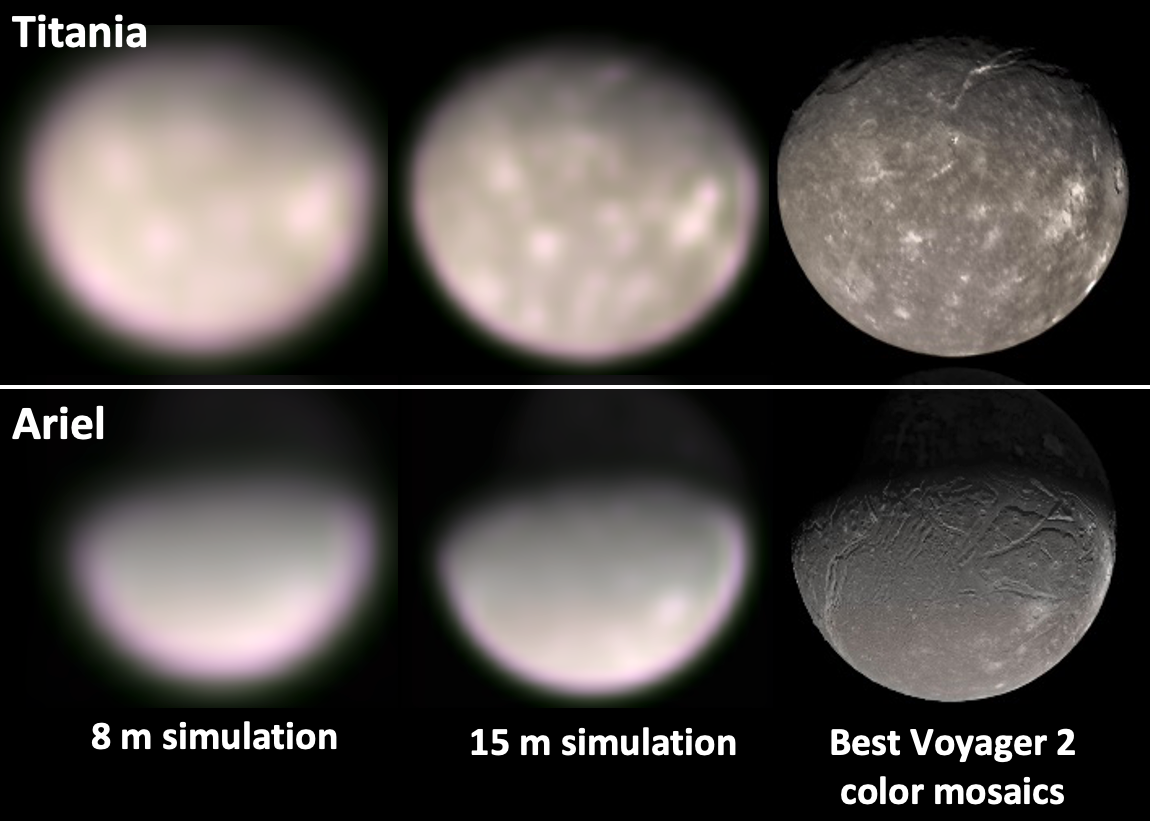}
	\caption{\textit{Resampled and real images of Titania and Ariel. Real images (right) are Voyager 2/ISS image mosaics (courtesy NASA/JPL/Caltech/USGS, reprocessed by \citet{stryk2008voyager}). Resampled images simulate what these moons would look like as seen by LUVOIR from the Earth-Sun L2 Lagrange point in 2040 with an 8 m (left) and 15 m (center) aperture. The simulated angular diameters are $\sim$0.08'' and $\sim$0.12'' for Ariel and Titania, respectively.}}\vspace{0.1 cm}
\end{figure} 

An orbiting spacecraft equipped with a magnetometer could search for induced magnetic fields emanating from briny oceans in the interiors of the classical moons. An orbiter could also search for plumes on the classical moons and other signs of recent endogenic geologic activity. An orbiter making multiple close flybys of Uranus' rings would dramatically improve our understanding of the ring moons and could investigate whether Mab is the source of the $\mu$-ring. An orbiter could spend time looking outward, making key observations of the distant irregular satellites, similar to Cassini’s observations of Saturn's irregular satellites \citep[]{denk2019studies}. A close pass of an irregular satellite inbound to Uranus, like Cassini's inbound flyby of Phoebe \citep[]{porco2005cassini}, would represent an unparalleled opportunity to investigate the nature and origin of these likely captured objects. Thus, new datasets collected by an orbiter are essential for improving our understanding of the icy residents of the Uranian system and determining whether Uranus' five classical moons are, or were, ocean worlds.
 
\section{Acknowledgments} 
This work was funded by NASA ROSES Solar System
Observations grant NNX17AG15G and Solar System Workings grant NHH18ZDA001N. This work was supported by the JPL Research and Technology Development Fund. Portions of this work were carried out at the Jet Propulsion Laboratory, California Institute of Technology, under contract to the National Aeronautics and Space Administration. We thank Roser Juanola-Parramon for providing the resampled Voyager 2 images of Ariel and Titania shown in Figure 7. We also thank Devon Burr and Michael Lucas for providing insightful feedback. 

\bibliography{references}{}

\begin{thebibliography}{}
\expandafter\ifx\csname natexlab\endcsname\relax\def\natexlab#1{#1}\fi
\providecommand{\url}[1]{\href{#1}{#1}}
\providecommand{\dodoi}[1]{doi:~\href{http://doi.org/#1}{\nolinkurl{#1}}}
\providecommand{\doeprint}[1]{\href{http://ascl.net/#1}{\nolinkurl{http://ascl.net/#1}}}
\providecommand{\doarXiv}[1]{\href{https://arxiv.org/abs/#1}{\nolinkurl{https://arxiv.org/abs/#1}}}

\bibitem[{Afanasiev {et~al.}(2014)Afanasiev, Rosenbush, \&
  Kiselev}]{afanasiev2014polarimetry}
Afanasiev, V., Rosenbush, V., \& Kiselev, N. 2014, Astrophysical Bulletin, 69,
  211

\bibitem[{Bauer {et~al.}(2002)Bauer, Roush, Geballe, Meech, Owen, Vacca,
  Rayner, \& Jim}]{bauer2002near}
Bauer, J.~M., Roush, T.~L., Geballe, T.~R., {et~al.} 2002, Icarus, 158, 178

\bibitem[{Beddingfield {et~al.}(2015)Beddingfield, Burr, \&
  Emery}]{beddingfield2015fault}
Beddingfield, C., Burr, D., \& Emery, J. 2015, Icarus, 247, 35

\bibitem[{Beddingfield {et~al.}(2020)Beddingfield, Li, Atreya, Beauchamp,
  Cohen, Fortney, Hammel, Hedman, Hofstadter, Rymer,
  {et~al.}}]{beddingfield2020exploration}
Beddingfield, C., Li, C., Atreya, S., {et~al.} 2020, arXiv preprint
  arXiv:2007.11063

\bibitem[{Beddingfield \& Cartwright(2020)}]{beddingfield2020hidden}
Beddingfield, C.~B., \& Cartwright, R.~J. 2020, Icarus, 113687

\bibitem[{Beddingfield \& Cartwright(2021)}]{beddingfield2021Arielcryo}
---. 2021, Under Review

\bibitem[{Bell~III \& McCord(1991)}]{bell1991search}
Bell~III, J., \& McCord, T. 1991, in Lunar and Planetary Science Conference
  Proceedings, Vol.~21, 473--489

\bibitem[{Brown \& Calvin(2000)}]{brown2000evidence}
Brown, M.~E., \& Calvin, W.~M. 2000, Science, 287, 107

\bibitem[{Brown \& Clark(1984)}]{brown1984surface}
Brown, R.~H., \& Clark, R.~N. 1984, Icarus, 58, 288

\bibitem[{Brown \& Cruikshank(1983)}]{brown1983uranian}
Brown, R.~H., \& Cruikshank, D.~P. 1983, Icarus, 55, 83

\bibitem[{Buratti \& Mosher(1991)}]{buratti1991comparative}
Buratti, B.~J., \& Mosher, J.~A. 1991, Icarus, 90, 1

\bibitem[{Capaccioni {et~al.}(2015)Capaccioni, Coradini, Filacchione, Erard,
  Arnold, Drossart, De~Sanctis, Bockelee-Morvan, Capria, Tosi,
  {et~al.}}]{capaccioni2015organic}
Capaccioni, F., Coradini, A., Filacchione, G., {et~al.} 2015, Science, 347

\bibitem[{Cartwright {et~al.}(2020{\natexlab{a}})Cartwright, Beddingfield,
  Showalter, Cruikshank, \& Nordheim}]{cartwright2020regolith}
Cartwright, R., Beddingfield, C., Showalter, M., Cruikshank, D., \& Nordheim,
  T. 2020{\natexlab{a}}, in Lunar and Planetary Science Conference No. 2326,
  1699

\bibitem[{Cartwright {et~al.}(2021)Cartwright, Nordheim, Grundy, DeColibus,
  Sory, Beddingfield, Leonard, Elder, Cochrane, Regoli, Atkinson, Holler,
  Cruikshank, \& Emery}]{cartwright2021arielCO2}
Cartwright, R., Nordheim, T., Grundy, W., {et~al.} 2021, in Lunar and Planetary
  Science Conference, Vol.~52

\bibitem[{Cartwright {et~al.}(2020{\natexlab{b}})Cartwright, Emery, Grundy,
  Cruikshank, Beddingfield, \& Pinilla-Alonso}]{cartwright2020probing}
Cartwright, R.~J., Emery, J.~P., Grundy, W.~M., {et~al.} 2020{\natexlab{b}},
  Icarus, 338, 113513

\bibitem[{Cartwright {et~al.}(2018)Cartwright, Emery, Pinilla-Alonso, Lucas,
  Rivkin, \& Trilling}]{cartwright2018red}
Cartwright, R.~J., Emery, J.~P., Pinilla-Alonso, N., {et~al.} 2018, Icarus,
  314, 210

\bibitem[{Cartwright {et~al.}(2015)Cartwright, Emery, Rivkin, Trilling, \&
  Pinilla-Alonso}]{cartwright2015distribution}
Cartwright, R.~J., Emery, J.~P., Rivkin, A.~S., Trilling, D.~E., \&
  Pinilla-Alonso, N. 2015, Icarus, 257, 428

\bibitem[{Cartwright {et~al.}(2019)Cartwright, Holler, Benecchi,
  Juanola-Parramon, Arney, Roberge, \& Hammel}]{cartwright2019exploring}
Cartwright, R.~J., Holler, B., Benecchi, S., {et~al.} 2019, arXiv preprint
  arXiv:1903.07691

\bibitem[{Cartwright {et~al.}(2020{\natexlab{c}})Cartwright, Beddingfield,
  Nordheim, Roser, Grundy, Hand, Emery, Cruikshank, \&
  Scipioni}]{cartwright2020evidence}
Cartwright, R.~J., Beddingfield, C.~B., Nordheim, T.~A., {et~al.}
  2020{\natexlab{c}}, The Astrophysical Journal Letters, 898, L22

\bibitem[{Clark \& Lucey(1984)}]{clark1984spectral}
Clark, R.~N., \& Lucey, P.~G. 1984, Journal of Geophysical Research: Solid
  Earth, 89, 6341

\bibitem[{Clark {et~al.}(2005)Clark, Brown, Jaumann, Cruikshank, Nelson,
  Buratti, McCord, Lunine, Baines, Bellucci, {et~al.}}]{clark2005compositional}
Clark, R.~N., Brown, R.~H., Jaumann, R., {et~al.} 2005, Nature, 435, 66

\bibitem[{Cochrane {et~al.}(2021)Cochrane, Nordheim, Vance, Styczinski,
  Soderlund, Elder, Leonard, Atkinson, Cartwright, Beddingfield, \&
  Regoli}]{cochrane2021induced}
Cochrane, C., Nordheim, T., Vance, S., {et~al.} 2021, 52

\bibitem[{Cohen {et~al.}(2020)Cohen, Beddingfield, Chancia, DiBraccio, Hedman,
  MacKenzie, Mauk, Sayanagi, Soderlund, Turtle, {et~al.}}]{cohen2020new}
Cohen, I., Beddingfield, C., Chancia, R., {et~al.} 2020, 51

\bibitem[{Cook {et~al.}(2018)Cook, Dalle~Ore, Protopapa, Binzel, Cartwright,
  Cruikshank, Earle, Grundy, Ennico, Howett, {et~al.}}]{cook2018composition}
Cook, J.~C., Dalle~Ore, C.~M., Protopapa, S., {et~al.} 2018, Icarus, 315, 30

\bibitem[{Croft(1987)}]{croft1987miranda}
Croft, S. 1987, in Lunar and Planetary Science Conference, Vol.~18

\bibitem[{Croft \& Soderblom(1991)}]{croft1991geology}
Croft, S., \& Soderblom, L. 1991, Uranus, 561

\bibitem[{Cruikshank {et~al.}(1977)Cruikshank, Pilcher, \&
  Morrison}]{cruikshank1977identification}
Cruikshank, D., Pilcher, C.~B., \& Morrison, D. 1977, The Astrophysical
  Journal, 217, 1006

\bibitem[{Cruikshank(1980)}]{cruikshank1980near}
Cruikshank, D.~P. 1980, Icarus, 41, 246

\bibitem[{Cruikshank \& Brown(1981)}]{cruikshank1981uranian}
Cruikshank, D.~P., \& Brown, R.~H. 1981, Icarus, 45, 607

\bibitem[{Cruikshank {et~al.}(2020)Cruikshank, Pendleton, \&
  Grundy}]{cruikshank2020organic}
Cruikshank, D.~P., Pendleton, Y.~J., \& Grundy, W.~M. 2020, Life, 10, 126

\bibitem[{Cruikshank {et~al.}(2019)Cruikshank, Umurhan, Beyer, Schmitt, Keane,
  Runyon, Atri, White, Matsuyama, Moore, {et~al.}}]{cruikshank2019recent}
Cruikshank, D.~P., Umurhan, O.~M., Beyer, R.~A., {et~al.} 2019, Icarus, 330,
  155

\bibitem[{{\'C}uk {et~al.}(2020){\'C}uk, El~Moutamid, \&
  Tiscareno}]{cuk2020dynamical}
{\'C}uk, M., El~Moutamid, M., \& Tiscareno, M.~S. 2020, The Planetary Science
  Journal, 1, 22

\bibitem[{Dalle~Ore {et~al.}(2019)Dalle~Ore, Cruikshank, Protopapa, Scipioni,
  McKinnon, Cook, Grundy, Schmitt, Stern, Moore, {et~al.}}]{dalle2019detection}
Dalle~Ore, C., Cruikshank, D., Protopapa, S., {et~al.} 2019, Science advances,
  5, eaav5731

\bibitem[{De~Pater {et~al.}(2006)De~Pater, Hammel, Gibbard, \&
  Showalter}]{de2006new}
De~Pater, I., Hammel, H.~B., Gibbard, S.~G., \& Showalter, M.~R. 2006, Science,
  312, 92

\bibitem[{DeColibus {et~al.}(2020)DeColibus, Chanover, \&
  Cartwright}]{decolibus2020investigating}
DeColibus, D., Chanover, N., \& Cartwright, R. 2020, AAS/Division for Planetary
  Sciences Meeting, 52, 215

\bibitem[{Denk \& Mottola(2019)}]{denk2019studies}
Denk, T., \& Mottola, S. 2019, Icarus, 322, 80

\bibitem[{Detre {et~al.}(2020)Detre, M{\"u}ller, Klaas, Marton, Linz, \&
  Balog}]{detre2020herschel}
Detre, {\"O}., M{\"u}ller, T., Klaas, U., {et~al.} 2020, Astronomy \&
  Astrophysics, 641, A76

\bibitem[{Emery {et~al.}(2005)Emery, Burr, Cruikshank, Brown, \&
  Dalton}]{emery2005near}
Emery, J., Burr, D., Cruikshank, D., Brown, R.~H., \& Dalton, J. 2005,
  Astronomy \& Astrophysics, 435, 353

\bibitem[{Fazio {et~al.}(2004)Fazio, Hora, Allen, Ashby, Barmby, Deutsch,
  Huang, Kleiner, Marengo, Megeath, {et~al.}}]{fazio2004infrared}
Fazio, G., Hora, J., Allen, L., {et~al.} 2004, The Astrophysical Journal
  Supplement Series, 154, 10

\bibitem[{Fletcher {et~al.}(2020)Fletcher, Helled, Roussos, Jones, Charnoz,
  Andr{\'e}, Andrews, Bannister, Bunce, Cavali{\'e},
  {et~al.}}]{fletcher2020ice}
Fletcher, L.~N., Helled, R., Roussos, E., {et~al.} 2020, Planetary and Space
  Science, 105030

\bibitem[{French {et~al.}(2017)French, Showalter, de~Pater, \&
  Lissauer}]{french2017orbital}
French, R.~S., Showalter, M.~R., de~Pater, I., \& Lissauer, J.~J. 2017, in
  AAS/Division for Planetary Sciences Meeting Abstracts\# 49, 214--19

\bibitem[{Gladman {et~al.}(2000)Gladman, Kavelaars, Holman, Petit, Scholl,
  Nicholson, \& Burns}]{gladman2000discovery}
Gladman, B., Kavelaars, J., Holman, M., {et~al.} 2000, Icarus, 147, 320

\bibitem[{Gladman {et~al.}(1998)Gladman, Nicholson, Burns, Kavelaars, Marsden,
  Williams, \& Offutt}]{gladman1998discovery}
Gladman, B.~J., Nicholson, P.~D., Burns, J.~A., {et~al.} 1998, Nature, 392, 897

\bibitem[{Gourgeot {et~al.}(2014)Gourgeot, Dumas, Merlin, Vernazza, \&
  Alvarez-Candal}]{gourgeot2014near}
Gourgeot, F., Dumas, C., Merlin, F., Vernazza, P., \& Alvarez-Candal, A. 2014,
  Astronomy \& Astrophysics, 562, A46

\bibitem[{Grav {et~al.}(2004)Grav, Holman, \& Fraser}]{grav2004photometry}
Grav, T., Holman, M.~J., \& Fraser, W.~C. 2004, The Astrophysical Journal
  Letters, 613, L77

\bibitem[{Graykowski \& Jewitt(2018)}]{graykowski2018colors}
Graykowski, A., \& Jewitt, D. 2018, The Astronomical Journal, 155, 184

\bibitem[{Greenberg(1991)}]{greenberg1991miranda}
Greenberg, R.~J. 1991, in Uranus (University of Arizona Press; Space Science
  Series), 693--735

\bibitem[{Grundy {et~al.}(2006)Grundy, Young, Spencer, Johnson, Young, \&
  Buie}]{grundy2006distributions}
Grundy, W., Young, L., Spencer, J., {et~al.} 2006, Icarus, 184, 543

\bibitem[{Grundy {et~al.}(2003)Grundy, Young, \& Young}]{grundy2003discovery}
Grundy, W., Young, L., \& Young, E. 2003, Icarus, 162, 222

\bibitem[{Grundy {et~al.}(2016)Grundy, Binzel, Buratti, Cook, Cruikshank,
  Dalle~Ore, Earle, Ennico, Howett, Lunsford, {et~al.}}]{grundy2016surface}
Grundy, W., Binzel, R., Buratti, B., {et~al.} 2016, Science, 351, aad9189

\bibitem[{Hammond \& Barr(2014)}]{hammond2014global}
Hammond, N.~P., \& Barr, A.~C. 2014, Geology, 42, 931

\bibitem[{Hanel {et~al.}(1986)Hanel, Conrath, Flasar, Kunde, Maguire, Pearl,
  Pirraglia, Samuelson, Cruikshank, Gautier, {et~al.}}]{hanel1986infrared}
Hanel, R., Conrath, B., Flasar, F., {et~al.} 1986, Science, 233, 70

\bibitem[{Hansen(1997)}]{hansen1997spectral}
Hansen, G.~B. 1997, Advances in Space Research, 20, 1613

\bibitem[{Helfenstein {et~al.}(1991)Helfenstein, Hillier, Weitz, \&
  Veverka}]{helfenstein1991oberon}
Helfenstein, P., Hillier, J., Weitz, C., \& Veverka, J. 1991, Icarus, 90, 14

\bibitem[{Helfenstein {et~al.}(1989)Helfenstein, Thomas, \&
  Veverka}]{helfenstein1989evidence}
Helfenstein, P., Thomas, P.~C., \& Veverka, J. 1989, Nature, 338, 324

\bibitem[{Hendrix {et~al.}(2019)Hendrix, Hurford, Barge, Bland, Bowman,
  Brinckerhoff, Buratti, Cable, Castillo-Rogez, Collins,
  {et~al.}}]{hendrix2019nasa}
Hendrix, A.~R., Hurford, T.~A., Barge, L.~M., {et~al.} 2019, Astrobiology, 19,
  1

\bibitem[{Hofstadter {et~al.}(2019)Hofstadter, Simon, Atreya, Banfield,
  Fortney, Hayes, Hedman, Hospodarsky, Mandt, Masters,
  {et~al.}}]{hofstadter2019uranus}
Hofstadter, M., Simon, A., Atreya, S., {et~al.} 2019, Planetary and Space
  Science, 177, 104680

\bibitem[{Hussmann {et~al.}(2015)Hussmann, Sotin, \&
  Lunine}]{hussmann2015interiors}
Hussmann, H., Sotin, C., \& Lunine, J. 2015, Treatise on Geophysics: Second
  Edition, 10, 605

\bibitem[{Irvine {et~al.}(2003)Irvine, Bergman, Lowe, Matthews, McGonagle,
  Nummelin, \& Owen}]{irvine2003hcn}
Irvine, W.~M., Bergman, P., Lowe, T.~B., {et~al.} 2003, Origins of Life and
  Evolution of the Biosphere, 33, 609

\bibitem[{Jacobson {et~al.}(1992)Jacobson, Campbell, Taylor, \&
  Synnott}]{jacobson1992masses}
Jacobson, R., Campbell, J., Taylor, A., \& Synnott, S. 1992, The Astronomical
  Journal, 103, 2068

\bibitem[{Jewitt \& Haghighipour(2007)}]{jewitt2007irregular}
Jewitt, D., \& Haghighipour, N. 2007, Annu. Rev. Astron. Astrophys., 45, 261

\bibitem[{Kargel(1995)}]{kargel1995cryovolcanism}
Kargel, J. 1995, in Comparative Planetology with an Earth Perspective
  (Springer), 101--113

\bibitem[{Karkoschka(2001)}]{karkoschka2001comprehensive}
Karkoschka, E. 2001, Icarus, 151, 51

\bibitem[{Kavelaars {et~al.}(2004)Kavelaars, Holman, Grav, Milisavljevic,
  Fraser, Gladman, Petit, Rousselot, Mousis, \&
  Nicholson}]{kavelaars2004discovery}
Kavelaars, J., Holman, M., Grav, T., {et~al.} 2004, Icarus, 169, 474

\bibitem[{Kirchoff \& Schenk(2009)}]{kirchoff2009crater}
Kirchoff, M.~R., \& Schenk, P. 2009, Icarus, 202, 656

\bibitem[{Kollmann {et~al.}(2020)Kollmann, Cohen, Allen, Clark, Roussos, Vines,
  Dietrich, Wicht, de~Pater, Runyon, {et~al.}}]{kollmann2020magnetospheric}
Kollmann, P., Cohen, I., Allen, R., {et~al.} 2020, Space Science Reviews, 216,
  1

\bibitem[{Leonard {et~al.}(2021)Leonard, Elder, Nordheim, Cartwright, Patthoff,
  Beddingfield, Tiscareno, Strange, \& Tibor}]{leonard2021umami}
Leonard, E., Elder, C., Nordheim, T., {et~al.} 2021

\bibitem[{Lewis(1972)}]{lewis1972metal}
Lewis, J.~S. 1972, Earth and Planetary Science Letters, 15, 286

\bibitem[{Maris {et~al.}(2007)Maris, Carraro, \& Parisi}]{maris2007light}
Maris, M., Carraro, G., \& Parisi, M.~G. 2007, Astronomy \& Astrophysics, 472,
  311

\bibitem[{McCord {et~al.}(1997)McCord, Carlson, Smythe, Hansen, Clark,
  Hibbitts, Fanale, Granahan, Segura, Matson, {et~al.}}]{mccord1997organics}
McCord, T.~a., Carlson, R., Smythe, W., {et~al.} 1997, Science, 278, 271

\bibitem[{McKinnon {et~al.}(1991)McKinnon, Chapman, \&
  Housen}]{mckinnon1991cratering}
McKinnon, W.~B., Chapman, C.~R., \& Housen, K.~R. 1991, Uranus, 629

\bibitem[{Mennella {et~al.}(2004)Mennella, Palumbo, \&
  Baratta}]{mennella2004formation}
Mennella, V., Palumbo, M., \& Baratta, G. 2004, The Astrophysical Journal, 615,
  1073

\bibitem[{Moore {et~al.}(2004)Moore, Schenk, Bruesch, Asphaug, \&
  McKinnon}]{moore2004large}
Moore, J.~M., Schenk, P.~M., Bruesch, L.~S., Asphaug, E., \& McKinnon, W.~B.
  2004, Icarus, 171, 421

\bibitem[{Nimmo \& Pappalardo(2016)}]{nimmo2016ocean}
Nimmo, F., \& Pappalardo, R. 2016, Journal of Geophysical Research: Planets,
  121, 1378

\bibitem[{Pappalardo {et~al.}(1997)Pappalardo, Reynolds, \&
  Greeley}]{pappalardo1997extensional}
Pappalardo, R.~T., Reynolds, S.~J., \& Greeley, R. 1997, Journal of Geophysical
  Research: Planets, 102, 13369

\bibitem[{Paradis {et~al.}(2019)Paradis, Moeckel, Tollefson, \&
  de~Pater}]{paradis2019photometry}
Paradis, S., Moeckel, C., Tollefson, J., \& de~Pater, I. 2019, The Astronomical
  Journal, 158, 178

\bibitem[{Peterson {et~al.}(2015)Peterson, Nimmo, \&
  Schenk}]{peterson2015elastic}
Peterson, G., Nimmo, F., \& Schenk, P. 2015, Icarus, 250, 116

\bibitem[{Plescia(1987)}]{plescia1987geological}
Plescia, J. 1987, Nature, 327, 201

\bibitem[{Porco {et~al.}(2005)Porco, Baker, Barbara, Beurle, Brahic, Burns,
  Charnoz, Cooper, Dawson, Del~Genio, {et~al.}}]{porco2005cassini}
Porco, C., Baker, E., Barbara, J., {et~al.} 2005, science, 307, 1237

\bibitem[{Protopapa {et~al.}(2020)Protopapa, Cook, Grundy, Cruikshank,
  Dalle~Ore, \& Beyer}]{protopapa2020charon}
Protopapa, S., Cook, J., Grundy, W., {et~al.} 2020, Pluto System after New
  Horizons, In Prep

\bibitem[{Raut {et~al.}(2012)Raut, Fulvio, Loeffler, \&
  Baragiola}]{raut2012radiation}
Raut, U., Fulvio, D., Loeffler, M., \& Baragiola, R. 2012, The Astrophysical
  Journal, 752, 159

\bibitem[{Rayner {et~al.}(2003)Rayner, Toomey, Onaka, Denault, Stahlberger,
  Vacca, Cushing, \& Wang}]{rayner2003spex}
Rayner, J., Toomey, D., Onaka, P., {et~al.} 2003, Publications of the
  Astronomical Society of the Pacific, 115, 362

\bibitem[{Roberge {et~al.}(2019)Roberge, Bolcar, \&
  France}]{roberge2019telling}
Roberge, A., Bolcar, M.~R., \& France, K.~C. 2019, in UV/Optical/IR Space
  Telescopes and Instruments: Innovative Technologies and Concepts IX, Vol.
  11115, International Society for Optics and Photonics, 111150O

\bibitem[{Romon {et~al.}(2001)Romon, De~Bergh, Barucci, Doressoundiram, Cuby,
  Le~Bras, Dout{\'e}, \& Schmitt}]{romon2001photometric}
Romon, J., De~Bergh, C., Barucci, M., {et~al.} 2001, Astronomy \& Astrophysics,
  376, 310

\bibitem[{Roush {et~al.}(1997)Roush, Noll, Cruikshank, \&
  Pendleton}]{roush1997uv}
Roush, T., Noll, K., Cruikshank, D., \& Pendleton, Y. 1997, in AAS/Division for
  Planetary Sciences Meeting Abstracts\# 29, 20--03

\bibitem[{Schenk(1991)}]{schenk1991fluid}
Schenk, P.~M. 1991, Journal of Geophysical Research: Solid Earth, 96, 1887

\bibitem[{Schenk \& Moore(2020)}]{schenk2020topography}
Schenk, P.~M., \& Moore, J.~M. 2020, Philosophical Transactions of the Royal
  Society A, 378, 20200102

\bibitem[{Sfair \& Winter(2012)}]{sfair2012role}
Sfair, R., \& Winter, S.~G. 2012, Astronomy \& Astrophysics, 543, A17

\bibitem[{Sheppard {et~al.}(2005)Sheppard, Jewitt, \&
  Kleyna}]{sheppard2005ultradeep}
Sheppard, S.~S., Jewitt, D., \& Kleyna, J. 2005, The Astronomical Journal, 129,
  518

\bibitem[{Showalter(2020)}]{showalter2020rings}
Showalter, M.~R. 2020, Philosophical Transactions of the Royal Society A, 378,
  20190482

\bibitem[{Showalter \& Lissauer(2006)}]{showalter2006second}
Showalter, M.~R., \& Lissauer, J.~J. 2006, Science, 311, 973

\bibitem[{Smith {et~al.}(1986)Smith, Soderblom, Beebe, Bliss, Boyce, Brahic,
  Briggs, Brown, Collins, Cook, {et~al.}}]{smith1986voyager}
Smith, B.~A., Soderblom, L., Beebe, R., {et~al.} 1986, Science, 233, 43

\bibitem[{Soifer {et~al.}(1981)Soifer, Neugebauer, \&
  Matthews}]{soifer1981near}
Soifer, B., Neugebauer, G., \& Matthews, K. 1981, Icarus, 45, 612

\bibitem[{Sori {et~al.}(2017)Sori, Bapst, Bramson, Byrne, \&
  Landis}]{sori2017wunda}
Sori, M.~M., Bapst, J., Bramson, A.~M., Byrne, S., \& Landis, M.~E. 2017,
  Icarus, 290, 1

\bibitem[{Spohn \& Schubert(2003)}]{spohn2003oceans}
Spohn, T., \& Schubert, G. 2003, Icarus, 161, 456

\bibitem[{Steinbr{\"u}gge {et~al.}(2019)Steinbr{\"u}gge, Steinke, Thor, Stark,
  \& Hussmann}]{steinbrugge2019measuring}
Steinbr{\"u}gge, G., Steinke, T., Thor, R., Stark, A., \& Hussmann, H. 2019,
  Geosciences, 9, 320

\bibitem[{Stryk \& Stooke(2008)}]{stryk2008voyager}
Stryk, T., \& Stooke, P. 2008, in Lunar and Planetary Science Conference,
  Vol.~39, 1362

\bibitem[{Tamayo {et~al.}(2013)Tamayo, Burns, \& Hamilton}]{tamayo2013chaotic}
Tamayo, D., Burns, J.~A., \& Hamilton, D.~P. 2013, Icarus, 226, 655

\bibitem[{Tittemore \& Wisdom(1990)}]{tittemore1990tidal}
Tittemore, W.~C., \& Wisdom, J. 1990, Icarus, 85, 394

\bibitem[{Verbiscer {et~al.}(2006)Verbiscer, Peterson, Skrutskie, Cushing,
  Helfenstein, Nelson, Smith, \& Wilson}]{verbiscer2006near}
Verbiscer, A.~J., Peterson, D.~E., Skrutskie, M.~F., {et~al.} 2006, Icarus,
  182, 211

\bibitem[{Waite~Jr {et~al.}(2009)Waite~Jr, Lewis, Magee, Lunine, McKinnon,
  Glein, Mousis, Young, Brockwell, Westlake, {et~al.}}]{waite2009liquid}
Waite~Jr, J.~H., Lewis, W., Magee, B., {et~al.} 2009, Nature, 460, 487

\bibitem[{Weiss {et~al.}(2021)Weiss, Biersteker, Colicci, Couch, Petropoulos,
  \& Balint}]{weiss2021searching}
Weiss, B., Biersteker, J., Colicci, V., {et~al.} 2021, in Lunar and Planetary
  Science Conference No. 2548, 2096

\bibitem[{Werner {et~al.}(2004)Werner, Roellig, Low, Rieke, Rieke, Hoffmann,
  Young, Houck, Brandl, Fazio, {et~al.}}]{werner2004spitzer}
Werner, M.~W., Roellig, T., Low, F., {et~al.} 2004, The Astrophysical Journal
  Supplement Series, 154, 1

\bibitem[{Wong {et~al.}(2020)Wong, Meech, Dickinson, Greathouse, Cartwright,
  Chanover, \& Tiscareno}]{wong2020transformative}
Wong, M.~H., Meech, K.~J., Dickinson, M., {et~al.} 2020, arXiv preprint
  arXiv:2009.08029

\bibitem[{Zahnle {et~al.}(2003)Zahnle, Schenk, Levison, \&
  Dones}]{zahnle2003cratering}
Zahnle, K., Schenk, P., Levison, H., \& Dones, L. 2003, Icarus, 163, 263

\end{thebibliography}
\bibliographystyle{aasjournal}



%

\end{document}